# Original Article

Running title: Automated and multi-parametric algorithm for meibography images

# An automated and multi-parametric algorithm for objective analysis of meibography images

Peng Xiao[1,#], Zhongzhou Luo[1,#], Yuqing Deng[1], Gengyuan Wang[1], and Jin Yuan[1,*]

[1]State Key Laboratory of Ophthalmology, Zhongshan Ophthalmic Center, Sun Yat-Sen University, Guangzhou 510060, China

[#]Authors contributed equally.

*Corresponding authors:

    Jin Yuan, MD, PhD

    Mailing address: Zhongshan Ophthalmic Centre, Sun Yat-Sen University,

    54 Xianlie Road, Guangzhou, China, 510060

    Tel.: (86)13825141659

    Fax: (8620) 87331550

    Email: yuanjincornea@126.com

# Abstract

**Bcakground**：Meibography is a non-contact imaging technique used by ophthalmologists and eye care practitioners to acquire information on the characteristics of meibomian glands. One of its most important applications is to assist in the evaluation and diagnosis of meibomian gland dysfunction (MGD). While artificial qualitative analysis of meibography images could lead to low repeatability and efficiency, automated and quantitative evaluation would greatly benefit the image analysis process. Moreover, since the morphology and function of meibomian glands varies at different MGD stages, multi-parametric analysis offering more comprehensive information could help in discovering subtle changes of meibomian glands during MGD progression. Therefore, automated and multi-parametric objective analysis of meibography images is highly demanded.

**Methods:** The algorithm is developed to perform multi-parametric analysis of meibography images with fully automatic and repeatable segmentation based on image contrast enhancement and noise reduction. The full architecture can be divided into three steps: (1) segmentation of the tarsal conjunctiva area as the region of interest (ROI); (2) segmentation and identification of glands within the ROI; and (3) quantitative multi-parametric analysis including newly defined gland diameter deformation index ($DI$), gland tortuosity index ($TI$), and glands signal index ($SI$). To evaluate the performance of the automated algorithm, the similarity index ($k$) and the segmentation error including the false positive rate ($r_P$) and the false negative rate ($r_N$) are calculated between the manually defined ground truth and the automatic segmentations of both the ROI and meibomian glands of 15 typical meibography images. The feasibility of the algorithm is demonstrated in analyzing typical meibograhy images.

**Results:** The results of the performance evaluation between the manually defined ground truth and the automatic segmentations are as following: for ROI segmentation, the similarity index $k = 0.94 \pm 0.02$, the false positive rate $r_P = 6.02 \pm 2.41\%$, and the false negative rate $r_N = 6.43 \pm 1.98\%$; for meibomian glands segmentation, the similarity index $k = 0.87 \pm 0.01$, the false positive rate $r_P = 4.35\% \pm 1.50\%$, and the false negative rate $r_N = 18.61\% \pm 1.54\%$. The algorithm has been successfully applied to process typical meibography images acquired from subjects at different meibomian gland healthy status, providing the glands area ratio $GA$, the gland length $L$, gland width $D$, gland diameter deformation index $DI$, gland tortuosity index $TI$ and glands signal index $SI$.

**Conclusions:** A fully automated algorithm has been developed showing high similarity with moderate segmentation errors for meibography image segmentation compared with the manual approach, offering multiple parameters to quantify the morphology and function of meibomian glands for objective evaluation of meibography image.

# 1. INTRODUCTION:

Meibomian glands have the functions of secreting various lipid components, forming the lipid layer, and preventing excessive tear evaporation, which are essential for maintaining ocular surface health and integrity (1). Functional disorders of the meibomian glands, known as meibomian gland dysfunction (MGD), are increasingly recognized as a high incidence disease (2-5). MGD is commonly characterized by terminal duct obstruction and/or abnormal glandular secretion, often leading to ocular surface epithelium damage, chronic blepharitis, and dry eye disease, which severely degrade life quality (6). While series of studies have been conducted on the prevalence of MGD, the reported results vary widely from 3.6% to 69.3% (7-8), which are mostly due to the lack of effective and unified diagnostic criteria[8]. Evaluation of the changes of meibomian glands is important for the diagnosis and management of MGD in clinical practice.

Besides traditional clinical evaluation methods combining clinical manifests for MGD diagnosis (9), imaging techniques like meibography (10-13), optical coherence tomography (14,15), and in vivo confocal microscopy (16,17) have also been developed and applied to image meibomian glands. As a non-contact imaging technique with infrared illumination, meibography acquires images of the meibomian glands' silhouette from entire human eyelids that better assist ophthalmologists in evaluating meibomian glands and making standardized diagnosis and treatment strategies (9-13).

With meibography images, morphological changes of the glands like dilation, distortion, shortening, and atrophy could be directly observed and assessed visually and qualitatively (6,9,18).

To aid ophthalmologists in conducting quantitative evaluation of meibomian glands with meibography images, various researches have been conducted to obtain detailed scale parameters and evaluate the reliability of these quantitative grading methods (10-13,19-28). The most commonly used scale parameter to classify the meibography images is the dropout area or meibomian gland area (10-13,19-23). With Image J or other image editing software, studies have shown that the dropout area or meibomian gland area can be quantitatively calculated, however, mostly in a subjective scheme as gland/gland-loss region are manually selected (10-13,19-21). Thus, these methods are relatively time-consuming and can lead to obvious intra- and inter-observer variability (13,24,25), which limit their use in clinical studies where a large number of images need to be analysed efficiently. To address these limits, objective analysis with automated algorithms is highly demanded. Nevertheless, only a few automated approaches have been demonstrated (22,23,25-28) due to the difficulties in the eyelid area and meibomian glands segmentation with meibography images.

While meibomian glands dropout exists mostly in the advanced stage of MGD, comparison of the histologic sections of normal and obstructed human meibomian glands have revealed that obstruction of the orifice in MGD could result in dilation of the gland acinus and central duct, increase of glands tortuosity, damage of the secretory meibocytes, and eventually atrophy of the meibomian glands (29,30). Therefore, except for the area ratio parameters, other quantitative parameters that can indicate detailed morphological changes of meibomian glands could potentially be used for MGD diagnosis and severity assessment (9). To date, only a handful of studies have reported the analysis of the gland length, width, or regularity in meibography images (12,13,25-28), in which some are subjectively characterized manually (12,13). A multi-parametric objective analysis of meibomian glands would offer doctors comprehensive information in MGD diagnosis and evaluation, especially in discovering subtle changes in preclinical and early-stage cases and monitoring MGD progression.

In this study, we developed an objective meibography image analysing algorithm that is capable to automatically perform segmentation of the eyelid area and all the single meibomian glands and provide multiple quantitative parameters for meibomain glands evaluation. Besides the traditional parameters like meibomian glands area ratio, gland length, and gland width, our algorithm introduced new morphological parameters, gland diameter deformation index and gland tortuosity index, to address the concerns in quantifying the local gland diameter variation and the degree of the curved and winding glands results from gland terminal duct obstruction, which mostly exists in intermediate stages of MGD progression (13,16,25,29,30). Moreover, since changes in the secretion, consistency and colour of meibum inside meibomian glands (12,29,31) could alter the signal intensity of glands in Meibography images, we developed a new parameter, gland signal index, to quantify the gland signal level in the meibography image, which can potentially be used to evaluate the meibum secretion activity. The performance of the automated algorithm is evaluated by analysing the similarity index and segmentation error between the manual and automatic segmentations. We demonstrated the feasibility of our algorithm in processing typical meibograhy images and compared the acquired gland parameters of difference mebography images and selected representative meibomian glands.

## 2. METHODS：

Our automated and multi-parametric algorithm for objective analysis of meibography images was developed based on Labview (Release 2016, National Instruments Corporation, US) with NI vision Development Module 2016 on a PC running Windows 10. The meibography images processed in this article were acquired using commercialized meibography Keratograph 5M (Oculus, Germany). The meibography acquires RGB infrared images and frames of the everted eyelid area with a resolution of 1088 x 512 pixel. Acquired images were converted to grayscale, exported and saved as bitmaps for analysing with our customized software. The algorithm provides fully automatic and repeatable analysis basically in three steps: (1) segment the tarsal conjunctiva area as the region of interest (ROI), (2) segment and identify meibomian glands within the ROI, and (3) quantitative multi-parametric analysis.

### 2.1 ROI segmentation

Fig. 1 shows the flow map of the processing steps of our algorithm for ROI segmentation with a sample meibography image (Fig.1 $I_G$). In brief, automatic extraction of the ROI is performed by removing the unwanted signals outside the tarsal conjunctiva area and fit the boundary of the ROI region. For this, the original grayscale meibography image (Fig.1 $I_G$) is firstly processed with a Prewitt operator to select the unwanted regions with eyelashes and highlight areas typically appear around the eyelash roots and the eyelid margin with large light reflections. The processed image is binarized and dilated to generate a mask image (Fig. 1 $I_M$), which is then subtracted from the original grayscale image (Fig. 1 $I_G$), results in image Fig. 1 $I_{GM}$. The resulting image undergoes median filtering with a circular structural element having 3 pixel diameter to remove random noises in the image. To enhance the image contrast and visibility of the glands, the Convolution-Highlight Details operator is applied to the image with a structural element size of 25x25 pixel. Glands edges and image details are further sharpened and outlined by a Laplacian filter with a structural element size of 29 X29 pixel, results in image Fig. 1 $I_{HD}$. The processed image is then binarized and binary inversed to highlight the glands surrounding backgrounds on the tarsal conjunctiva area (Fig. 1 $I_{BI}$). As the former subtracted regions are inversed to be positive, the mask image is subtracted again from the binary inversed image Fig. 1 $I_{BI}$, result in image Fig. 1 $I_{BIM}$. To eliminate image speckles and small artefacts, Fig. 1 $I_{BIM}$ is eroded by a structural element with a size of 5x5, which also shatters connected small irregularities. All signal blocks with a size less than 190 pixels, which are empirically found, are then cleared. This to remove those small irregularities that mostly do not belong to the everted eyelid. The image is then dilated back by a structural element with a size of 5x5, and a border rejecting operator is applied to remove large irregularities connected with the image borders (Fig. 1 $I_{RSO}$).

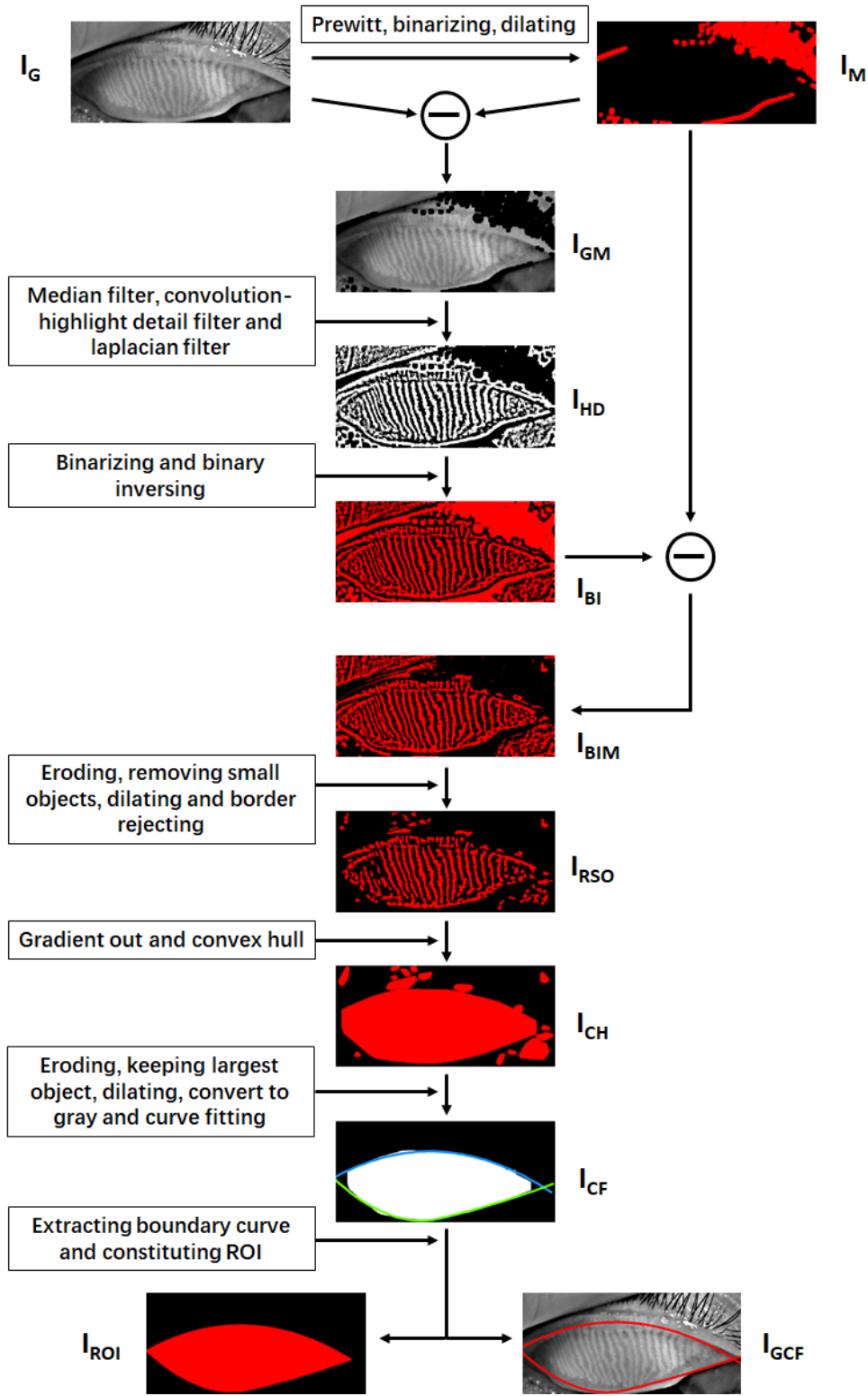

**Fig. 1.** Flow map showing the algorithm steps for ROI segmentation.

As most of the information of the tarsal conjunctiva area is kept, a gradient out operator with a circular structural element of 8 pixel diameter is employed to connect the signal blocks followed by a convex hull operator that using graham scan algorithm to fit a closing shape of the tarsal conjunctiva area (Fig. 1 $I_{CH}$). To get rid of those remaining surrounding lumps, the image is therefore eroded, withhold only the largest block, and dilated back, results in a single block fitting the eyelid region (Fig. 1 $I_{CF}$). To define the final ROI, Fig. 1 $I_{CF}$ is converted into grayscale, and the contour of the final block is used to fit the final boundaries of the ROI. For the upper boundary, longitudinal top-down scans are performed to find the first signal rising points, which are then connected and fitted by a $2^{nd}$ degree one-dimensional B-spline of the centripetal parameter (blue line in Fig. 1 $I_{CF}$). For the lower boundary, the optimal fitting with a $2^{nd}$ degree one-dimensional B-spline is performed to the line connecting the first signal rising points of bottom-up longitudinal scans in the image, (green line in Fig. 1 $I_{CF}$). The final ROI is acquired with the closing area formed by the fitted upper and lower boundaries (Fig. 1 $I_{ROI}$). The original meibography image with the ROI boundary (red line in Fig. 1 $I_{GCF}$), within which the gland segmentation and quantitative analysis would be performed, is also shown in Fig. 1 $I_{GCF}$.

**2.2 Segmentation and identification of glands within the ROI**

Fig. 2 shows the main steps of the algorithm for meibomian glands segmentation. The original grayscale meibography image (Fig. 2 $I_G$) is firstly processed by a median filter, a Convolution-Highlight Details operator, and a Laplacian filter sequentially, as aforementioned in the ROI segmentation process, which damp the random noises and highlight the glands' visibility (Fig. 2 $I_{GHD}$). The resulting image is then multiplied by the binary mask of the ROI and binarized (Fig. 2 $I_{GE}$). To remove the remaining image speckles and also smooth the shape of the glands, the binarized image is convoluted by a median filter with a circular structural element of 5 pixel diameter. Since the ROI segmentation is usually overestimated resulting in some detected objects belongs to the edge of the everted eyelid, an orientation operator is applied to remove objects with the main angle lower than 45° or higher than 135°, determined empirically, as the eyelid edges are typically horizontal while glands mostly have a vertical orientation. Fig. 2 $I_{GL}$ is the image contains all the detected signal from the glands for glands area ratio analysis.

Since those small meibomian glands on the nasal and temporal sides are mostly irregular and are usually out of focus and unevenly illuminated during meibography imaging, segmented meibomian glands (Fig. 2 $I_{GL}$) on the nasal and temporal sides are typically broken into pieces. Thus, to further perform quantitative analysis of every single gland, these broken glands are excluded by eliminating objects containing less than 1400 pixels, result in Fig. 2 $I_{MG}$. While in some cases segmented glands could be connected (Highlighted in green and yellow in Fig. 2 $I_{MG}$), affecting the subsequent analysis, these connected glands are detected by calculating the number of horizontal segments $N_h$ and vertical segments $N_v$ of the single objects in Fig. 2 $I_{MG}$ and extracting the ones with $N_h>350$ or $N_v>200$ (Fig. 2 $I_{CG}$). The extracted connected glands images (Fig. 2 $I_{CG}$) are sent to the glands fragmentation algorithm.

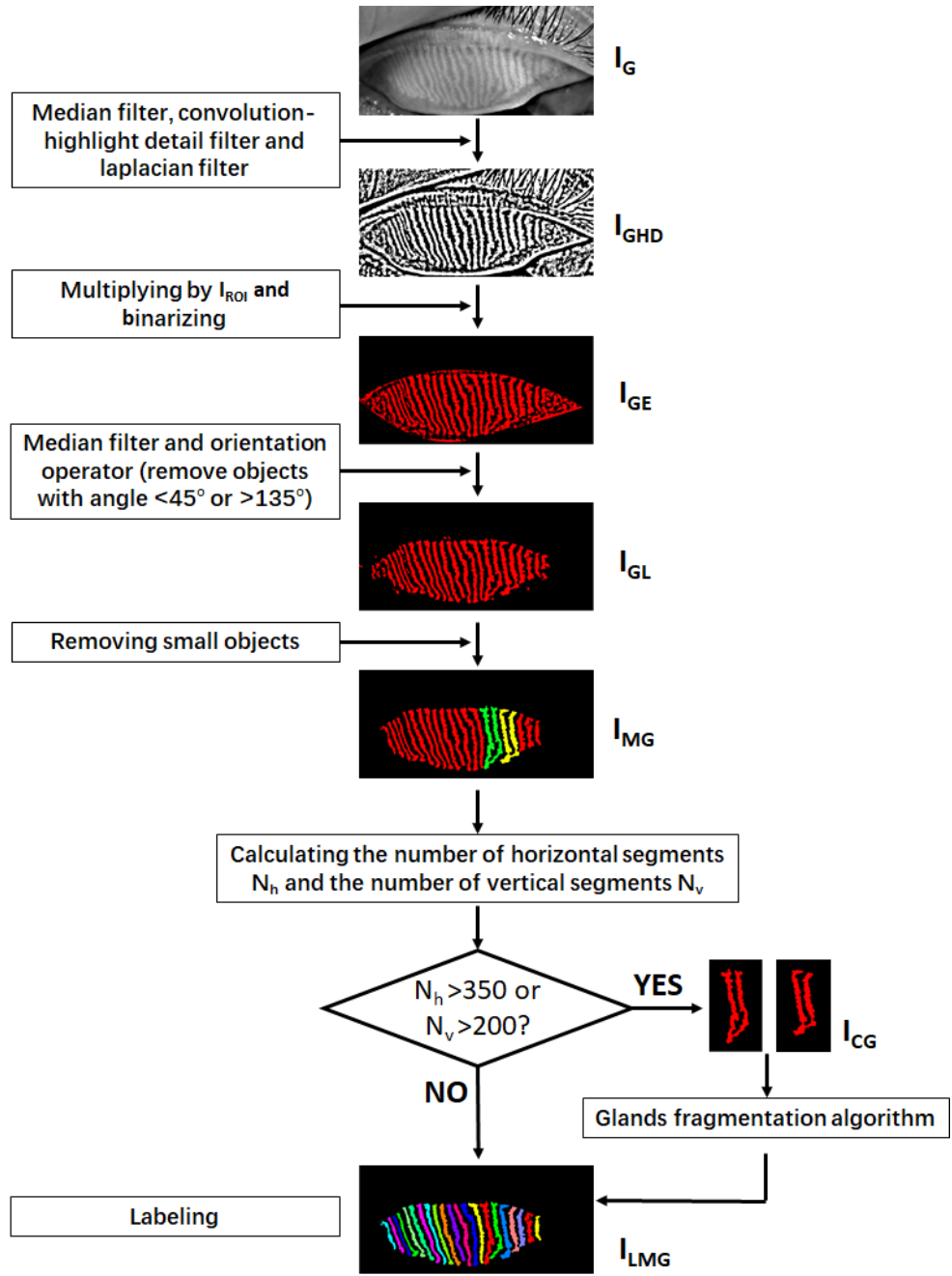

**Fig. 2.** Flow map showing the algorithm steps for meibomian glands segmentation. The glands fragmentation algorithm is depicted in Fig. 3.

As shown in Fig. 3, the glands fragmentation algorithm is developed to automatically separate those connected glands. The extracted connected glands image (Fig. 3 $I_{CG}$) is continuously eroded with a structural element size of 1×3 pixel until the detecting objects number N exceeds 1 (Fig. 3 $I_{ED}$). The eroded image is then binary inversed and skeletonized to generate a skeleton image (Fig. 3 $I_{IS}$), which is then subtracted from the connected glands image (Fig. 3 $I_{CG}$). This results in the separation of the connected glands (Fig. 3 $I_{GS}$). However, as seen in the zoom-in area of Fig. 3 $I_{GS}$, although separated, the single glands are disrupted by some skeleton branches. Before sending the separated disrupted glands for repairing, the number of horizontal segments $N_h$ and vertical segments $N_v$ of the single objects in Fig. 3 $I_{GS}$ are calculated to validate if there are still objects of connected glands that exist. If yes, it goes back to the eroding process to further separate them, otherwise, the separated single disrupted glands (Fig. 3 $I_{SGS}$) proceed to the repair process. By applying a gradient out operator with a circular structural element of 3 pixel diameter, following by a fill hole operator, the disrupted single glands are fixed while expended out by one pixel (Fig. 3 $I_{GO}$). A gradient in the operator with a circular structural element of 3 pixel diameter is then applied to form the contours of the single glands (Fig. 3 $I_{GC}$). The repaired glands (Fig. 3 $I_{RG}$) are achieved by subtracting the gland contours from the expended glands. The final separated glands are shown in Fig. 3 $I_{SG}$. By combing the final separated glands and those detected single glands (Red glands in Fig. 2 $I_{MG}$), all the intact glands are identified and labelled for further quantitative analysis.

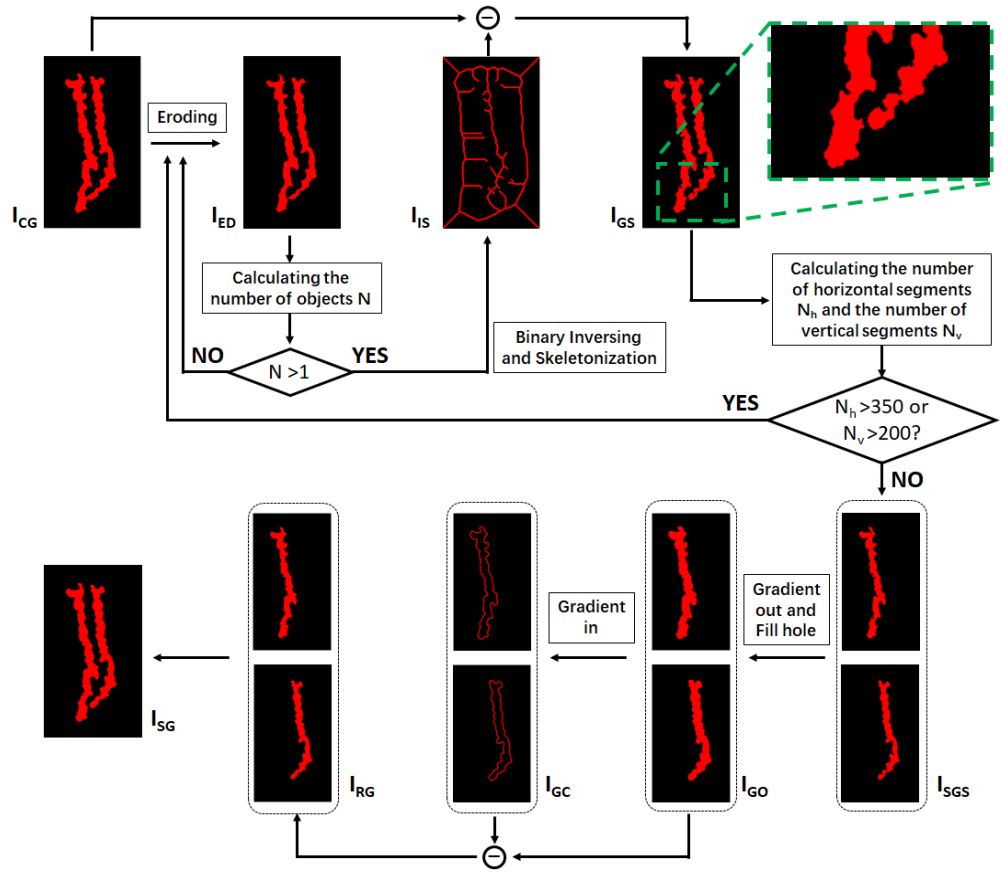

**Fig. 3.** Flow map of the glands fragmentation algorithm to separate connected glands.

**2.3 Quantitative parameters analysis**

With the segmented ROI and labelled intact meibomian glands, quantitative parameters including glands area ratio, gland length, gland width, gland diameter deformation index, gland tortuosity index, and glands signal index are defined and computed.

The glands area ratio $GA$ is calculated by:

$$GA = \frac{N_{gl}}{N_{ROI}} \times 100\% \tag{1}$$

in which $N_{gl}$ is the number of pixels occupied by all the detected signals from the glands (Fig. 2 $I_{GL}$) and $N_{ROI}$ is the number of pixels occupied by the segmented ROI area (Fig. 1 $I_{ROI}$).

Morphological parameters analysis of the length, width, diameter deformation index, and tortuosity index are performed on each segmented and labelled meibomian glands (Fig. 2 $I_{LMG}$). The single gland length $L$ is defined by:

$$L = R \times l_{MN} \tag{2}$$

in which $l_{MN}$ is the length (in pixels) of the gland central line between endpoints $M$ and $N$ (Fig. 4(a)), $R$ is the digital resolution of the meibography. By imaging a diffuse reflectance grid distortion target (#62-949, Edmund Optics), as shown in Fig. 4(b), $R$ is calibrated to be $0.03 mm/pixel$ in our case with images captured by OCULUS Keratograph 5M. The gland width $D$ is defined by the averaged width across a single gland:

$$D = R \times \frac{1}{n}\sum_{i=1}^{n} d_i \tag{3}$$

in which $d_i$ is the width (in pixels) of the gland at every 3 pixel points on the gland central line and is calculated by the pixel length of the perpendicular line of the local tangent line within the intersection points $A, B$ on the gland edge contour (Fig. 4(a)). To address the diameter variations like the uneven dilation and discontinuous atrophy of a gland, which are related to the status of meibomian gland duct and acinus[37], gland diameter deformation index $DI$ is introduced by:

$$DI = \sigma_d = \sqrt{\frac{1}{n}\sum_{i=1}^{n}(R \times d_i - D)^2} \tag{4}$$

which is actually the standard deviation $\sigma_d$ of the local gland widths within a single gland. To quantify the degree of the gland curving and hairpin-loop-like winding changes, which mostly result from gland terminal duct obstruction, gland tortuosity index $TI$ is defined as:

$$TI = \frac{l_{MN}}{l'_{MN}} \times \frac{1}{n}\sum_{i=1}^{n} k_i = \frac{l_{MN}}{l'_{MN}} \times \frac{1}{n}\sum_{i=1}^{n} \left|\frac{\Delta \alpha_i}{R \times \Delta s_i}\right| \tag{5}$$

in which $l_{MN}$ is the length (in pixels) of the gland central line between endpoints $M$ and $N$, $l'_{MN}$ is the length (in pixels) of the straight line between endpoints $M$ and $N$, $k_i$ is the local curvature of the gland central line at every 3 pixel points, $\Delta \alpha_i$ is the angularity between tangent lines at point $i$ and $i-1$, and $\Delta s_i$ is the arc length (in pixels) between point $i$ and $i-1$ (Fig. 4(a)). To have an overall evaluation of all the glands detected in the tarsal conjunctiva, averaged length $\bar{L}$, width $\bar{D}$, diameter deformation index $\overline{DI}$, and tortuosity index $\overline{TI}$ of all the detected meibomian glands are computed for each meibography image.

As meibomian gland functional and structural disorders affect the meibum secretion activity, different meibum consistency and colour directly alter the reflection of the illumination light, result in changes in the signal intensity of glands in meibography images. To quantify the gland signal level in the meibography image, which can potentially be used to evaluate the meibum secretion, the glands signal index is defined by:

$$SI = lg\frac{GREY_i}{GREY_0} \tag{6}$$

in which $GREY_i$ is the averaged image gray value of the segmented intact glands (Fig. 2 I$_{LMG}$) in raw meibography image (Fig. 2 I$_G$), and $GREY_0$ is the averaged image gray value of the non-gland area (Inversed area of Fig. 2 I$_{GE}$) within the RIO area in raw meibography image.

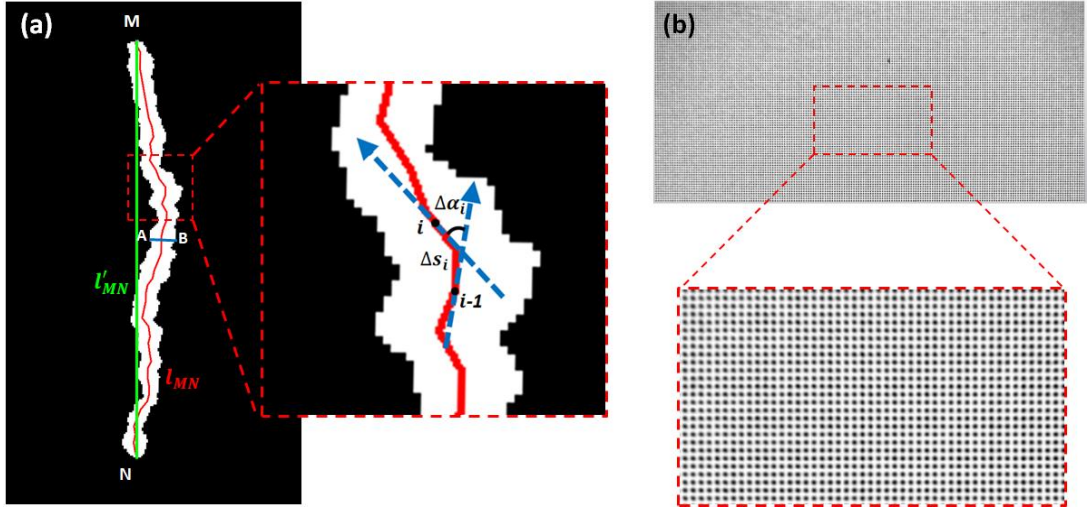

**Fig. 4.** (a) Schematic illustrating the elements for morphological parameter calculation of single meibomian gland; (b) Meibography imaging of a diffuse reflectance grid distortion target for digital resolution calibration of the system.

## 3. VALIDATION AND RESULTS

To validate the performance of our automated and multi-parametric algorithm, 15 meibography images were acquired with meibography Keratograph 5M (Oculus, Germany) from subjects at different meibomian gland healthy status evaluated and agreed by two professional ophthalmologists. Informed consent was obtained for the subjects. The experimental procedures adhered to the tenets of the Declaration of Helsinki (1983) and were approved by the Institutional Review Board of Zhongshan Ophthalmic Centre, Sun Yat-sen University, China (protocol number: 2019KYPJ110). The ground truth of the ROI and meibomian glands segmentations were defined by the mean manually delineated ROI and glands' boundaries by the two ophthalmologists. Manual segmentation costs around 8 minutes for one meibography image. The meibography images were also automatically and objectively processed by the algorithm for ROI and glands' segmentation as well as multiple parameters calculation. The automated processing lasted around 15 seconds for each meibography image.

### 3.1 Segmentation performance validation

To evaluate the segmentation performance of our automated algorithm, the similarity index ($k$) and the segmentation error (32-35) were calculated between the manually defined ground truth segmentation and the automatic segmentation of both the ROI and meibomian glands. The similarity index ($k$), also known as the kappa coefficient, measures the similarity of the two segmentations defined by the ratio of the area of their intersection divided by the area of their union (Equation 7). It ranges from 0 for segmentations that have no overlap to 1 for segmentations that are identical. For the segmentation error, we calculated the false positive rate ($r_P$) as well as the false negative rate ($r_N$). The false positive rate ($r_P$) was depicted as the probability of the non-ROI/non-gland area included in the automatic segmentation results, and the false negative rate ($r_N$) vice versa. The false positive rate ($r_P$) and the false negative rate ($r_N$) were calculated by diving the area of the incorrectly segmented non-ROI/non-gland area and ROI/gland area by the manually segmented ROI/gland area as shown in equation 8 and 9 respectively.

$$k(S_m, S_a) = \frac{2|S_m \cap S_a|}{|S_m| + |S_a|} \tag{7}$$

$$r_P = \frac{|S_a| - |S_m \cap S_a|}{|S_m|} \times 100\% \tag{8}$$

$$r_N = \frac{|S_m| - |S_m \cap S_a|}{|S_m|} \times 100\% \tag{9}$$

in which $S_m$ is the manually segmented ROI/gland area and $S_a$ is the automatic segmented ROI/gland area.

Fig. 5 illustrates the contours of the segmented ROI (Fig. 5(a)) and meibomian glands (Fig. 5(b)) of a representative meiborgraphy image. The contours of the mean manually segmented ROI and meibomian glands are depicted in yellow while the algorithm automatically segmented contours are depicted in red. Table 1 shows the averaged similarity index and segmentation error between the manual and automatic segmentation of the ROI and meibomian glands for the 15 meibography images. For ROI segmentation: the similarity index $k = 0.94 \pm 0.02$, the false positive rate $r_P = 6.02 \pm 2.41\%$, and the false negative rate $r_N = 6.43 \pm 1.98\%$. For meibomian glands segmentation: the similarity index $k = 0.87 \pm 0.01$, the false positive rate $r_P = 4.35\% \pm 1.50\%$, and the false negative rate $r_N = 18.61\% \pm 1.54\%$.

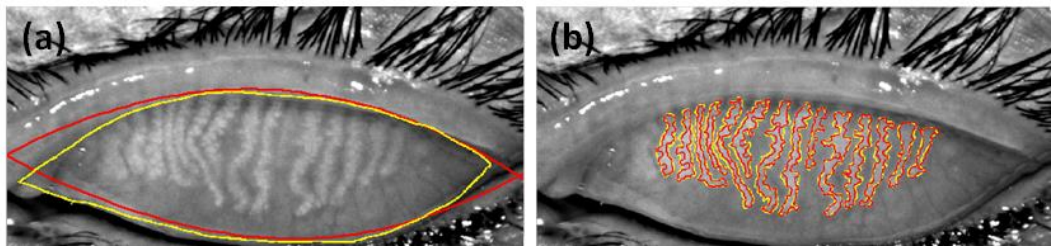

**Fig. 5.** Illustrations of the contours of the (a) ROI and (b) meibomian glands segmented manually (yellow) and automatically (red) of a representative meiborgraphy image.

**Table 1.** Averaged similarity index and segmentation error between the manual segmentation and automatic segmentation of the ROI and meibomian glands.

| | Similarity index $k$ | Segmentation error | |
|---|---|---|---|
| | | False positive rate $r_P(\%)$ | False negative rate $r_N(\%)$ |
| **ROI Segmentation** | 0.94±0.02 | 6.02±2.41 | 6.43±1.98 |
| **Glands Segmentation** | 0.87±0.01 | 4.35±1.50 | 18.61±1.54 |

Data are shown as mean±standard deviation. ROI: region of interest.

### 3.2 Automatic segmentation and multi-parametric results

Fig. 6 shows the segmentation results of the five representative meibography images acquired from subjects with different health status of meibomian glands. Images are displayed in Fig. 6(a) to Fig. 6 (e) based on their current stages of MGD progression from healthy to severe evaluated and agreed by two professional ophthalmologists. The left column shows the original meibography images with the red curves compassing the automatically segmented ROIs. The right column shows the segmented and labelled single meibomian glands for further quantitative analysis. The automatic calculated quantitative parameters, including glands area ratio *GA*, glands signal index *SI*, as well as averaged gland length $\bar{L}$, width $\bar{D}$, diameter deformation index $\overline{DI}$, and tortuosity index $\overline{TI}$, are shown in Table 2. From our preliminary results, with the MGD severity progressed from healthy to severe stages, the measured glands area ratio *GA* decreases from 39.84% to 19.34%; the averaged gland length $\bar{L}$ also shows a downtrend. Nevertheless, the averaged gland width $\bar{D}$, diameter deformation index $\overline{DI}$, and tortuosity index $\overline{TI}$ all raise in meibography images of intermediate MGD and then decline back in severe MGD with serious gland atrophy occurred. The glands signal index *SI* varies since the secretion, density, and colour of meibum inside the meibomian glands diverse in different cases.

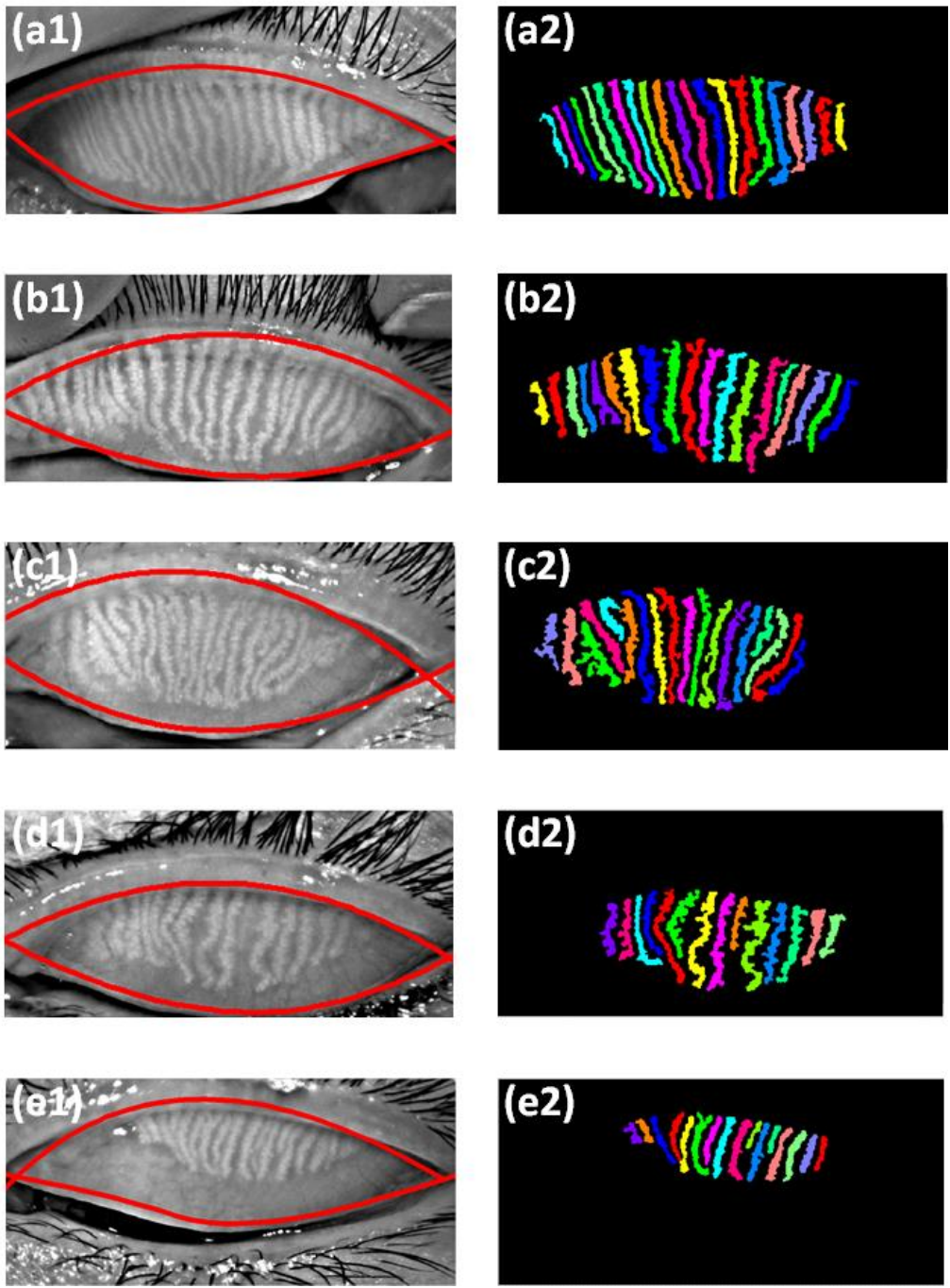

**Fig. 6.** The automatic ROI (left column) and meibomian glands (right column) segmentation results of the five representative meibography images with different healthy status of meibomian glands. The display sequence from (a) to (e) is based on their current stages of MGD progression from healthy to severe.

**Table 2.** Multi-parametric analysis results of the five representative meibography images with different healthy status of meibomian glands.

|   | GA (%) | $\bar{L}$ (mm) | $\bar{D}$ (mm) | $\overline{DI}$ | $\overline{TI}$ | SI |
|---|---|---|---|---|---|---|
| **a** | 39.84 | 5.62±1.25 | 0.35±0.08 | 0.10±0.04 | 0.18±0.11 | 13.91 |
| **b** | 36.73 | 5.04±1.20 | 0.42±0.07 | 0.12±0.03 | 0.26±0.08 | 20.27 |
| **c** | 31.57 | 5.05±1.26 | 0.42±0.10 | 0.15±0.07 | 0.30±0.26 | 19.53 |
| **d** | 29.20 | 4.55±1.25 | 0.42±0.08 | 0.15±0.04 | 0.40±0.21 | 18.15 |
| **e** | 19.34 | 2.62±0.76 | 0.32±0.06 | 0.09±0.04 | 0.18±0.09 | 24.97 |

Data are shown as mean±standard deviation. GA: glands area ratio; L: gland length; D: gland width; DI: gland diameter deformation index; TI: gland tortuosity index; SI: glands signal index.

### 3.3 Comparison of representative glands

To better demonstrate the correspondence of the gland morphological parameters to the detailed gland characteristics, especially the newly defined gland diameter deformation index $DI$ and tortuosity index $TI$, we selected several representative glands from the processed meibography images and compared the calculated morphological parameters and their appearance characteristics as shown in Table 3. It is obvious that the less uniform the gland diameter along its length, the higher the gland diameter deformation index $DI$, which is directly related to the uneven dilation and discontinuous atrophy of the meibomian gland duct and acinus (29,30). For the tortuosity index $TI$, the larger the degree of the gland curving and winding, the higher the tortuosity index $TI$, which reflects the macro tortuous appearance of meibomian gland typically appears in intermediate MGD (25,29).

Table 3. Morphological parameters of representative meibomian glands.

| Representative glands | 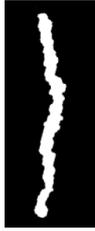 | 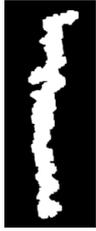 | 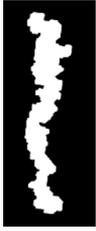 | 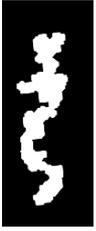 | 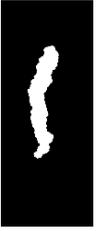 |
|---|---|---|---|---|---|
| L (mm) | 6.55 | 6.51 | 6.44 | 6.15 | 3.06 |
| D (mm) | 0.35 | 0.48 | 0.54 | 0.58 | 0.34 |
| DI | 0.07 | 0.13 | 0.18 | 0.25 | 0.09 |
| TI | 0.13 | 0.31 | 0.41 | 1.12 | 0.20 |

L: gland length; D: gland width; DI: gland diameter deformation index; TI: gland tortuosity index.

## 4. DISCUSSION AND CONCLUSIONS

In this study, we reported a fully automated algorithm for objective segmentation and multi-parametric quantitative analysis of meibography images. The automatic segmentations have shown high similarity with moderate segmentation errors compared with manual segmentations by professional ophthalmologists. Traditional parameters, such as glands area ratio, gland length and width, as well as newly defined parameters, including gland diameter deformation index, gland tortuosity index, and glands signal index, have been achieved to demonstrate a more comprehensive analysis of meibography images.

While various researches have conducted quantitative analysis of meibography images with everted eyelid area and meibomian glands segmentations, most of the studies implemented subjective methods to manually select the eyelid region and/or segment the gland/gland-loss area with commercialized image editing software like Image J (12,13,24) or self-developed semi-automatic algorithms (21), which usually have relatively low reliability and efficiency (24,25). To date, some automated methods have been developed to perform objective analysis of meibography images with only a few researches have demonstrated multi-parametric automated analysis. Arita et al. proposed image enhancement techniques to isolate meibomian glands but computing only the glands area ratio (22). Celik et al. introduced automated meibomian gland segmentation using Gabor wavelets filtering, extracting orientation, width, length, and curvature of segmented glands, nevertheless, it used a standardized elliptical area for the ROI segmentation without adapting it to the real shape of each meibography image (23). Koh et al. demonstrated an automated method to extract gland morphological features like gland length and width and trained a linear classifier for meibography image classification, however, with manual selection of the analysis eyelid region (26). Koprowski et al. demonstrated a fully automated approach that differentiates between gland, intergland and gland drop-out area to analyse the meibography images by glands area ratio together with detailed parameters quantifying the gland branches (27). Llorens-Quintana et al. described an algorithm that is capable of segment glands and tarsal conjunctiva area automatically, providing glands area ratio, fitted ellipse for gland length and width analysis, and a newly defined gland irregularity parameter (28). Compared with those pre-existed approaches, our newly developed automated and multi-parametric algorithm identifies the ROI contours and segment meibomian glands based on image contrast enhancement and noise reduction, sub-algorithm has also been implemented to separate connected glands while not affecting its original appearance for further morphological analysis. Besides traditional parameters, we proposed, for the first time to the best of our knowledge, to objectively quantify the gland diameter variations, the gland tortuosity as well as the glands signal level, thus offering more comprehensive information which is likely to be advantageous for detecting local and subtle changes of meibomian glands.

From the segmentation performance evaluation results, our automatic algorithm showed a relatively high false negative rate for meibomian glands segmentation compared with the mean manual segmentation. This is due to the fact that manual segmentation trends to include the gaps between gland acinus as the gland area while automatic algorithms are more sensitive to image contrast and can detect and exclude these gap areas. Compared with former studies, the glands area ratio calculated by our algorithm is relatively smaller (12). This is because we exclude the inter-gland area from calculating the glands area ratio as our method can precisely distinguish meibomian glands and inter-gland area, while most of the former studies have included it. The quality of the meibography images acquired for algorithm processing needs to be guaranteed. Images with low contrast, inhomogeneous illumination, and poor resolution could introduce extra error during ROI and meibomian gland segmentations, affecting the reliability of the final quantitative analysis results. Currently, our algorithm is adapted to the images acquired with meibography Keratograph 5M (Oculus, Germany) only as most of the processing parameters are selected empirically, its university for other types of meibograpy instruments can be adjusted.

Since our automated and multi-parametric algorithm offers more morphological as well as functional parameters for meibomian glands analysis and our preliminary results show its potential in detecting subtle variations in meibomian glands, large scale clinical studies will need to be conducted to comprehensively characterize the relations between these quantitative parameters and the clinical manifestation of meibomian glands in different pathological stages of diseases. Moreover, our current algorithm takes account only the analysis of upper lid meibography images. Since everting the lower eyelid is not as easy as the upper eyelid where a larger tarsal plate exists, the meibography image of the lower eyelid typically shows unevenly focused eyelid with partially exposed meibomian glands, thus, automatic segmentations of the ROI and meibomian glands in the lower lid remain challenging. Future work will focus on the lower lid meibography image analysis since combining the evaluation of both upper and lower lid will offer better clinical diagnostic performance.

**Acknowledgments**

Funding: This research is supported by the National Key R&D Program of China (2017YFC0112400) and the National Natural Science Foundation of China (81901788).

**Footnote**

*Conflicts of interest:* All authors have completed the ICMJE uniform disclosure form. The authors have no conflicts of interest to declare.

*Ethical Statement:* The experimental procedures adhered to the tenets of the Declaration of Helsinki (1983) and were approved by the Institutional Review Board of Zhongshan Ophthalmic Centre, Sun Yat-sen University, China (protocol number: 2019KYPJ110). Informed consent was obtained from all the subjects.